# An integrated electro-optically tunable multi-channel interference cavity laser

**Junxia Zhou,**[a] **Yiran Zhu,**[b] **Botao Fu,**[a] **Jinming Chen,**[a] **Huiting Song,**[a] **Zhihao Zhang,**[a] **Jianping Yu,**[a] **Jian Liu,**[b] **Min Wang,**[a] **Jia Qi,**[a, *] **Ya Cheng,**[a, b, c, e, f, g, *]

[a] The Extreme Optoelectromechanics Laboratory (XXL), School of Physics and Electronic Science, East China Normal University, Shanghai 200241, China
[b] State Key Laboratory of Precision Spectroscopy, East China Normal University, Shanghai 200062, China
[c] State Key Laboratory of High Field Laser Physics and CAS Center for Excellence in Ultra-Intense Laser Science, Shanghai Institute of Optics and Fine Mechanics (SIOM), Chinese Academy of Sciences (CAS), Shanghai 201800, China
[d] Center of Materials Science and Optoelectronics Engineering, University of Chinese Academy of Sciences, Beijing 100049, China
[e] Hefei National Laboratory, Hefei 230088, China
[f] Collaborative Innovation Center of Extreme Optics, Shanxi University, Taiyuan 030006, China
[g] Collaborative Innovation Center of Light Manipulations and Applications, Shandong Normal University, Jinan 250358, China

*Jia Qi, jqi@phy.ecnu.edu.cn; Ya Cheng, ya.cheng@siom.ac.cn

**Abstract**: We demonstrated a continuously tunable laser system by butt coupling a reflective semiconductor optical amplifier (RSOA) chip with a thin-film lithium niobate (TFLN) based multi-channel interference (MCI) cavity chip. This hybrid integrated lasers allows for fine-tuning of the laser wavelength from 1538 nm to 1560 nm with a resolution of 0.014 nm and a side-mode suppression ratio (SMSR) exceeding 30 dB. The MCI cavity chip is fabricated using the photolithography assisted chemo-mechanical etching (PLACE) technique. The developed laser has an output power of approximately 10 μW, which can be further amplified to 70 mW using a commercial erbium-doped fiber amplifier (EDFA) without significant broadening of the laser linewidth.

**Keywords**: Tunable laser, multi-channel interference, photonic integrated circuit (PIC), lithium niobite on insulator (LNOI)

## 1 Introduction

There is an urgent need for chip-level and chip-integrated lasers capable of continuous tuning over a wide wavelength range in the fields of optical communication, sensing, and measurement[1-5]. Currently, most high-quality tunable lasers rely on embedded gratings for mode selection to achieve wide continuous tuning of laser wavelengths. These lasers are exemplified by the distributed feedback (DFB) laser array and the distributed Bragg reflector (DBR) laser[6-10]. To reduce the complexity and cost of chip fabrication, alternative wavelength tuning strategies have been proposed, including Vernier microring resonators[11-15], etching grooves[16-19], and V-coupled cavities[20-23]. However, these strategies cannot simultaneously fulfill the requirements of high power, wide tuning range, and continuous tuning.

In 2015, Chen et al[24]. proposed employing a multi-channel interference (MCI) cavity for wavelength tuning, and demonstrated the fabrication of an InP-based monolithic tunable laser with a tuning resolution of 0.4 nm. The MCI cavity[25, 26] design gets ride of the need for grating fabrication, which simplifies the production process. Moreover, each interference arm can be individually phase-adjusted, which minimizes the influence of random initial phase on the stability of systems.

Thin film Lithium niobate (TFLN) is regarded as an ideal material for cavity fabrication owing to its exceptional nonlinear optical properties, ultra-wide optical and electrical bandwidths, and remarkably high electro-optical modulation rates and efficiency[27-31]. Meanwhile, advancements in device fabrication technology[32-38] for TFLN have been noteworthy in recent years. Consequently, hybrid integrated lasers, which involve integrating reflective semiconductor optical amplifiers (RSOAs) with TFLN external cavities, are being increasingly considered as a promising solution for on-chip laser applications.

In this paper, we demonstrate a hybrid integrated tunable laser comprising an MCI external cavity chip fabricated on TFLN and an RSOA chip. The laser exhibits a tuning range from 1538 nm to 1560 nm, a tuning resolution of 0.014 nm, and a side-mode-suppression-ratio (SMSR) exceeding 30 dB. The output power of the hybrid integrated tunable laser is amplified from 10 μW to 70 mW using a commercial Erbium-doped fiber amplifier (EDFA), while the linewidth is roughly unchanged after amplification.

## 2 Results

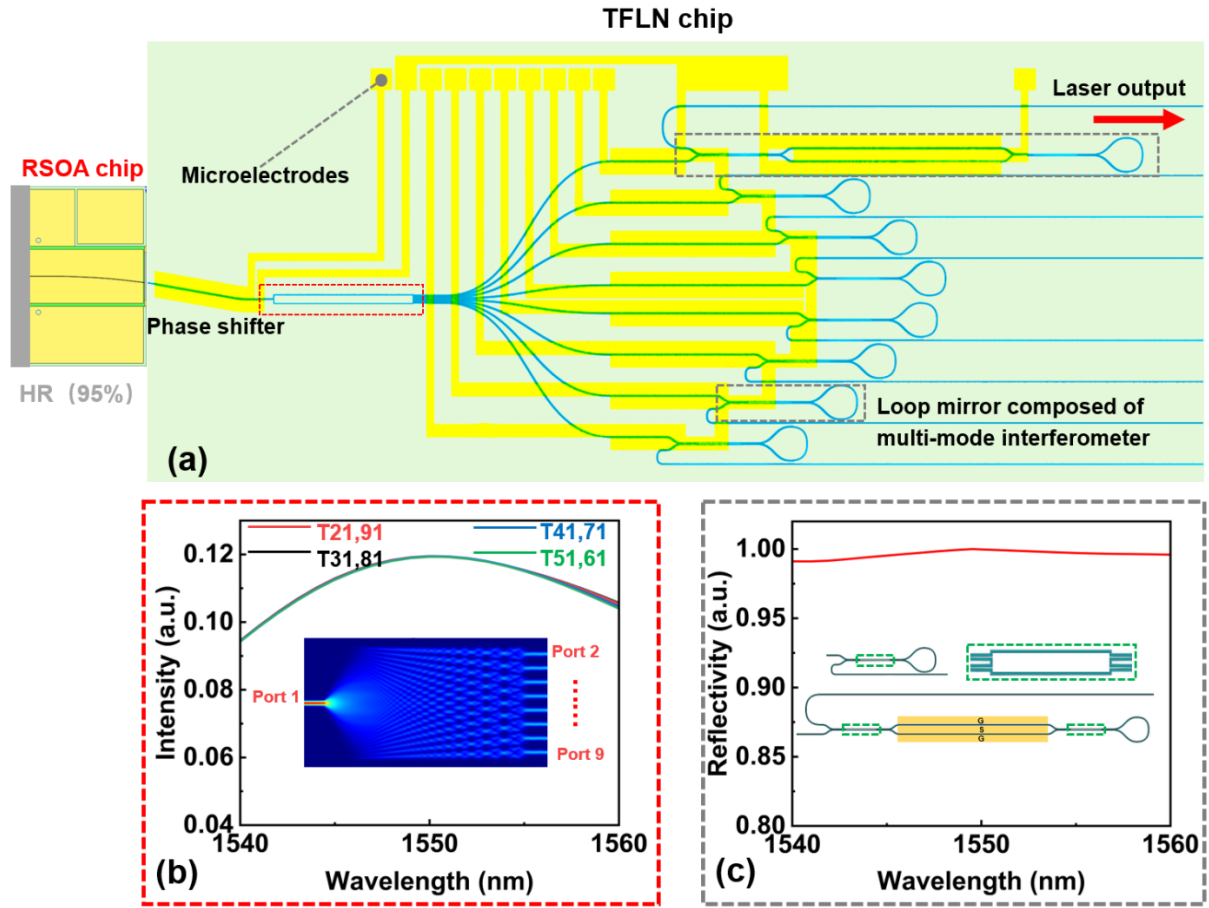

**Figure 1.** (a) Schematic view of the hybrid integrated tunable laser, which is composed of a InP-based RSOA chip and a TFLN-based MCI external cavity chip. (b) 1×8 MMI output power as a function of wavelength. The inset shows the field distributions simulation of a 1×8 MMI. (c) sagnac loop reflectors reflectivity as function of wavelength. The inset shows the schematics of the sagnac loop reflector and the tunable sagnac loop reflector.

The design of the hybrid integrated tunable laser is illustrated in Figure 1(a). The laser consists of a commercial InP-based RSOA chip (Anritsu, AE5T315BY20P) which provides C-band single-angled facet gain, and a TFLN-based MCI external cavity chip which provides optical feedback and consequently wavelength tuning. The RSOA chip incorporates a geometric design aimed at minimizing reflections from its front surface. This design involves curving the ridged waveguide at a 5° angle, effectively eliminating back-reflections that could otherwise introduce undesired feedback into the laser cavity. The output aperture width of RSOA is 3 μm. The Fabry–Perot (FP) cavity of the hybrid integrated tunable laser is defined between a 95% reflective coating in the RSOA chip and 8 sagnac loop reflectors (SLRs) in the MCI external cavity chip. The average cavity length is 16 mm, and the corresponding free spectral range (FSR) is 0.07 nm.

The MCI external cavity is crucial in achieving wide and continuous tuning of the laser wavelength. It has a 10° tilted input waveguide to match the output of the RSOA which can realize high-efficient coupling, which is followed by a common phase shifter, 1×8 multimode interferometers (MMI) splitter, 8 independent phase shifters with different arm lengths, 8 SLRs with one being a TSLR. The TSLR differs from the other SLRs in that it includes an MZI and a laser output waveguide. By adjusting the voltage applied to the MZI, the maximum output of the current wavelength can be achieved.

The 1×8 MMI splitter is designed using the Finite Difference Time Domain (FDTD) method to achieve 8 uniform splitting over a wide wavelength range. The obtained optimal geometric parameters are as followings: the top width of the input waveguide and the output waveguide is 1 μm, the top width of the interference region is 100 μm, and the length is 1680 μm. Figure 1(b) illustrates the simulation result of the 8-channel uniform splitting effect for the designed 1×8MMI.

The designs of 8 SLRs including one TSLR are similar to our previous work[39, 40], with one notable modification to increase the bandwidth: the direct couplers (DCs) in these sections have been substituted with 2×2 MMIs. Figure 1(c) illustrates the schematics of these sections and the simulated reflectivity of these SLRs which is over 98% from 1540 nm to 1560 nm.

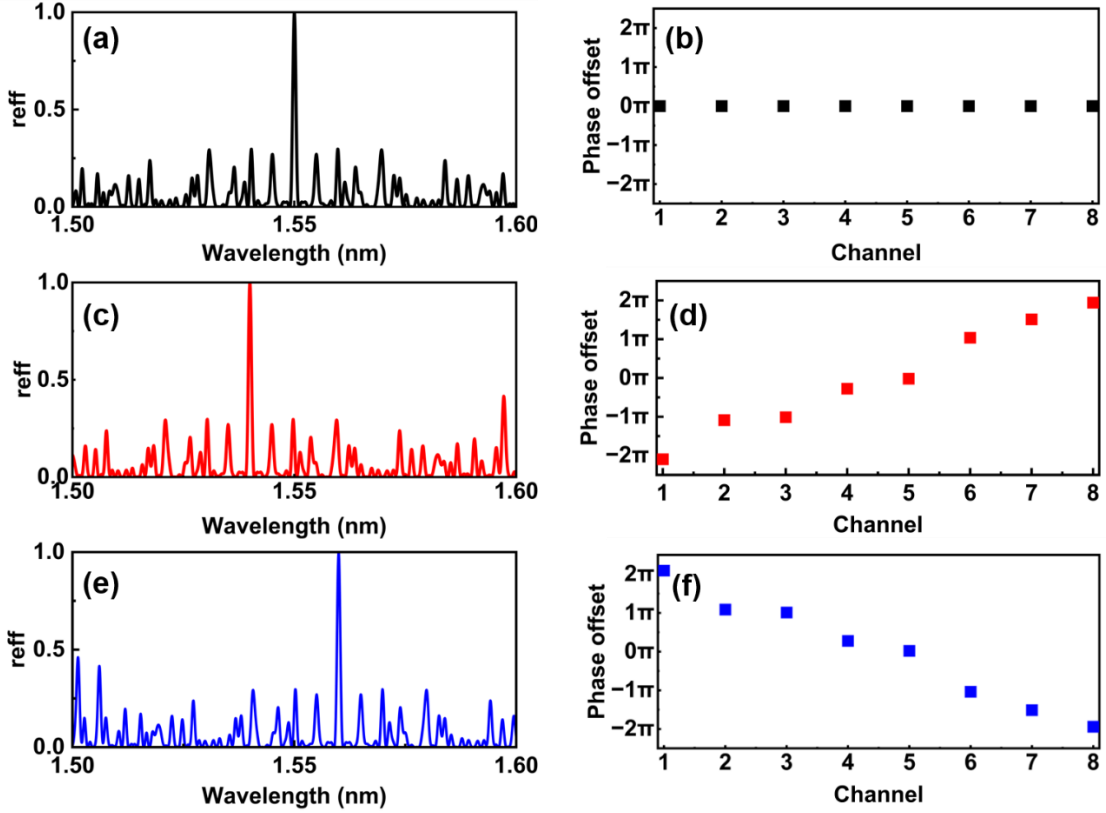

**Figure 2.** The relationship between the 8 independent phase shifts and the main reflection peak.

The aggregated complex reflection coefficient which the MCI external cavity provided can be expressed as：

$$\tilde{r}(\lambda) = \frac{1}{8} r \sum_i e^{j[2\tilde{\beta}(\bar{L}+\Delta L_i)+\phi_i]}$$

where $r$ is the reflection coefficient of single SLR, $\bar{L}$ is the average cavity length, $\Delta L_i$ is the length difference between the arm length of the $i_{th}$ independent phase shifter and $i_{th+1}$ independent phase shifter, $\phi_i$ is the $i_{th}$ independent phase shifter introduced by electro-optic modulation, $\tilde{\beta} = \beta - \frac{1}{2}j\alpha$ is the complex propagation constant of the guided mode, $\beta$ is the real constant of the guided mode and $\alpha$ is the average propagation loss of 8 paths.

$\Delta L_i$ is the key to achieving high SMSR in the reflection spectrum, which means that the arm lengths of the 8 independent phase shiftes need to be carefully optimized. After calculation (see supplementary materials for the process), the arm lengths of the 8 independent phase shifters were finally determined to be 3.3693 mm, 3.6530 mm, 3.8707 mm, 4.2404 mm, 4.3442 mm, 4.1257 mm, 3.8087 mm, 3.6126 mm. Figure 2 shows the simulated reflectance spectrum of the 8-channel interference at these length parameters. In this design, the central wavelength of the main reflection peak is 1550nm. At this wavelength, the additional electrical phase shifts of all eight independent phase shifters are assumed to be zero. When the main reflection peak shifts from the current wavelength to a specific wavelength, the 8 independent phase shifters will produce corresponding phase shifts. Similarly, by applying voltage to these independent phase shifters, they can generate corresponding phase shifts, allowing the main reflection peak to switch from the current wavelength to the specific wavelength.

Specifically, when the laser longitudinal mode is tuned by one FSR, the $i_{th}$ phase change is:

$$\delta\phi_i = 2\pi + 2\pi\frac{\Delta L_i}{\bar{L}} + \Delta\phi_i$$

Therefore, moving the $i_{th}$ independent phase shifter by $\Delta\phi_i = -2\pi\frac{\Delta L_i}{\bar{L}}$ can switch the optimal aggregation point of the complex reflection coefficient to the adjacent longitudinal mode, thereby achieving wavelength switching with tuning accuracy of FSR.

The common phase shifter is designed for fine-tuning of the laser wavelength. By varying the voltage applied to the common phase shifter, the overall cavity length of the laser can be continuously adjusted, thereby achieving precise wavelength tuning through slight changes in the cavity longitudinal mode.

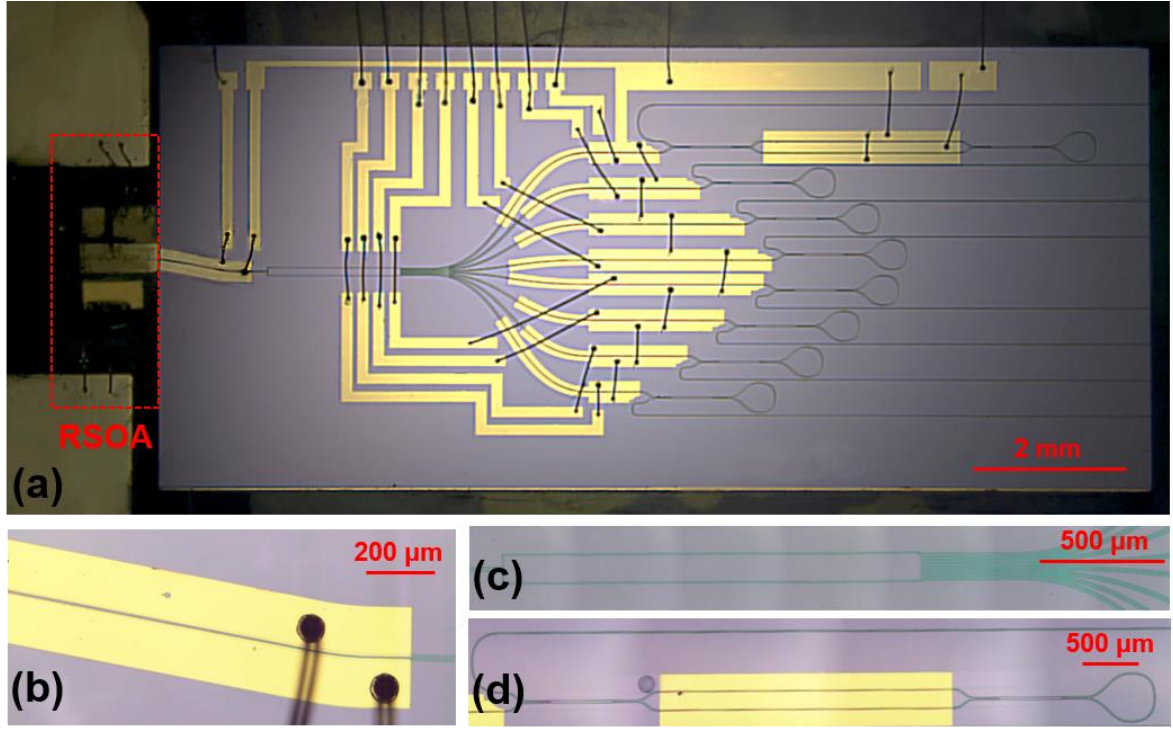

**Figure 3.** (a) optical micrograph of the hybrid integrated tunable laser, (b) Close-up optical micrograph of the 10° inclined waveguide. (c) Close-up optical micrograph of the 1x8 MMI. (d) Close-up optical micrograph of the TSLR.

The hybrid integrated tunable laser is processed according to the mentioned design. Figure 3(a) shows an optical micrograph of the prepared tunable laser. The majority on the right side of the image is the manufactured MCI external cavity chip which includes a 10° inclined waveguide with a common phase shifter, a 1x8 MMI beam splitter, 8 independent phase shifters (3.3693 mm, 3.6530 mm, 3.8707 mm, 4.2404 mm, 4.3442 mm, 4.1257 mm, 3.8087 mm, and 3.6126 mm from top to bottom, respectively), 8 SLRs with one TSLR. The MCI external cavity chip is fabricated on a 500-nm-thick X-cut TFLN substrate (500 nm TFLN/4.7 μm SiO2/500 μm Si) using the femtosecond laser-assisted chemical mechanical polishing (PLACE) method. More details about the PLACE method can be found in our previous work[41-43]. Figures 3(b)-(d) respectively show the microscopic details of the common phase shifter, 1×8MMI and TSLR in the MCI external cavity chip. The 1x8 MMI in the MCI chip has been proven to exhibit uniform 8-path splitting characteristics from 1538 nm to 1560 nm, and the SLRs in the MCI chip have been tested and found to have more than 90% reflectivity at these wavelengths.

The external voltage source, using the PCB board as a medium, applies voltage to the microelectrodes at various locations on the chip, achieving phase adjustment of the common phase shifter, 8 independent phase shifters, and the MZI in the TSLR. By butting the RSOA chip with the MCI external cavity the fabrication of the hybrid integrated tunable laser is finally realized.

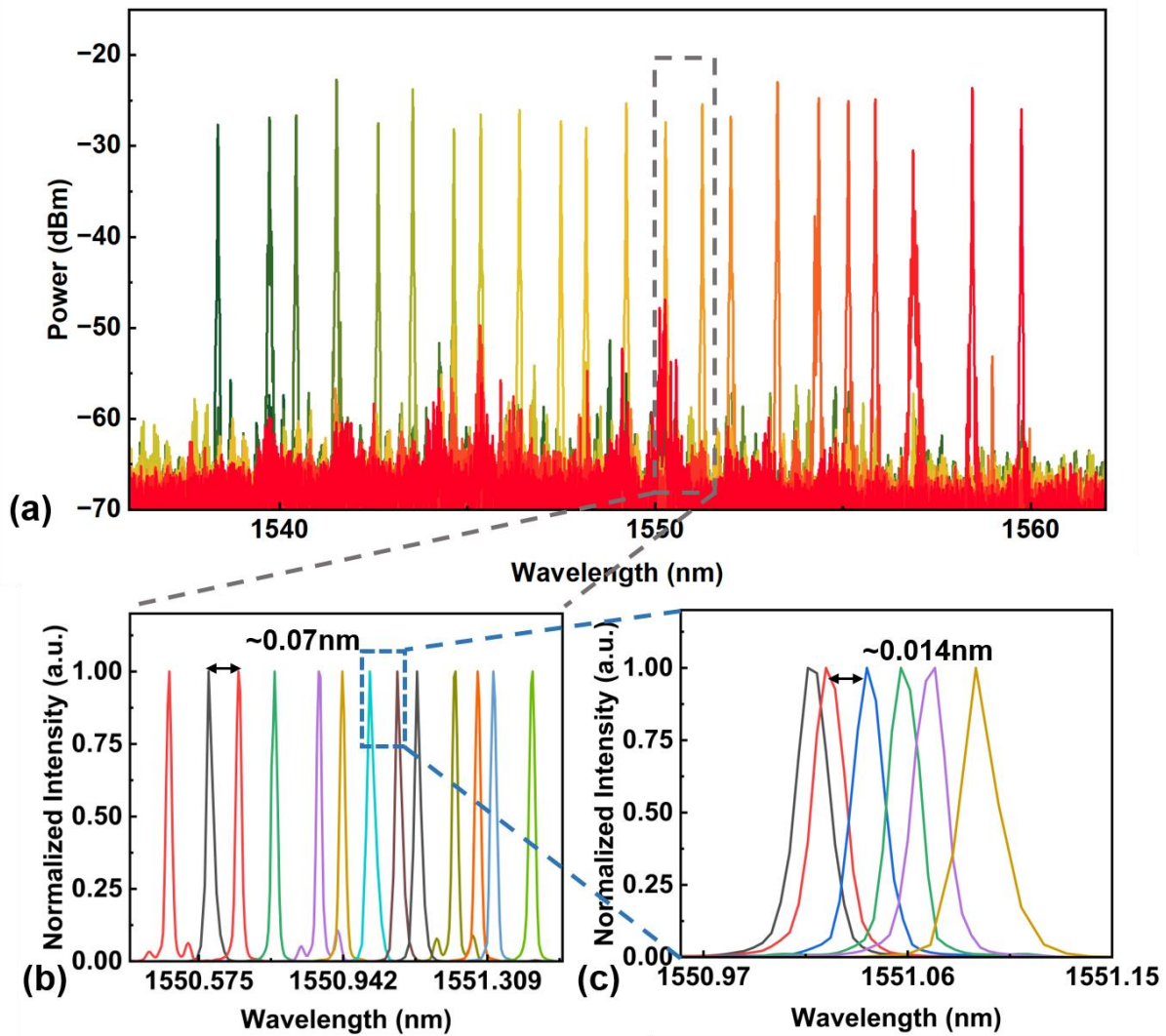

**Figure 4.** (a) Superimposed spectra of coarse tuning with a resolution of 1 nm. (b) Superimposed spectra of coarse tuning with a resolution of 0.07 nm. (c) Superimposed spectra of fine tuning with a resolution of 0.014 nm..

An optical spectrum analyzer (OSA: AQ6370D, Yokogawa) with a wavelength resolution of 20 pm is connected to the output of the MCI chip to demonstrate the wavelength tuning of the hybrid integrated tunable laser. Due to the impracticality of directly conducting high-precision wavelength tuning throughout the entire 1538nm to 1560nm wavelength range, which would result in a large amount of data for testing and demonstration, we demonstrate the wavelength tuning process through the following incremental steps：

Firstly, wavelength tuning is carried out with a resolution of 1 nm to characterize the overall tuning range of the tunable laser. As shown in Figure 4(a), by applying specific voltages to the eight independent phase shifters to generate specific phase shifts, the laser wavelength is tuned from 1538 nm to 1560 nm, achieving a wavelength tuning range of 22 nm.

Secondly, we select two longitudinal modes from Figure 4(a) (ranging from 1550.5 nm to 1551.425 nm) and conduct fine-tuning on eight independent phase shifters to achieve the laser spectrum shown in Figure 4(b). The resolution of the laser wavelength tuning is equivalent to the FSR of the laser, which is 0.07 nm, as previously mentioned.

Thirdly, by adjusting the phase of the common phase shifter, the overall cavity length of the laser is altered, thereby achieving continuous tuning of the laser wavelength. Two longitudinal modes are selected from Figure 4(b) (ranging from 1551.016 nm to 1551.09 nm). Within this range, by adjusting the phase of the common phase shifter, the spectrum shown in Figure 4(c) is obtained. The fine-tuning resolution of the laser wavelength is 0.014 nm (limited by the precision of the OSA device).

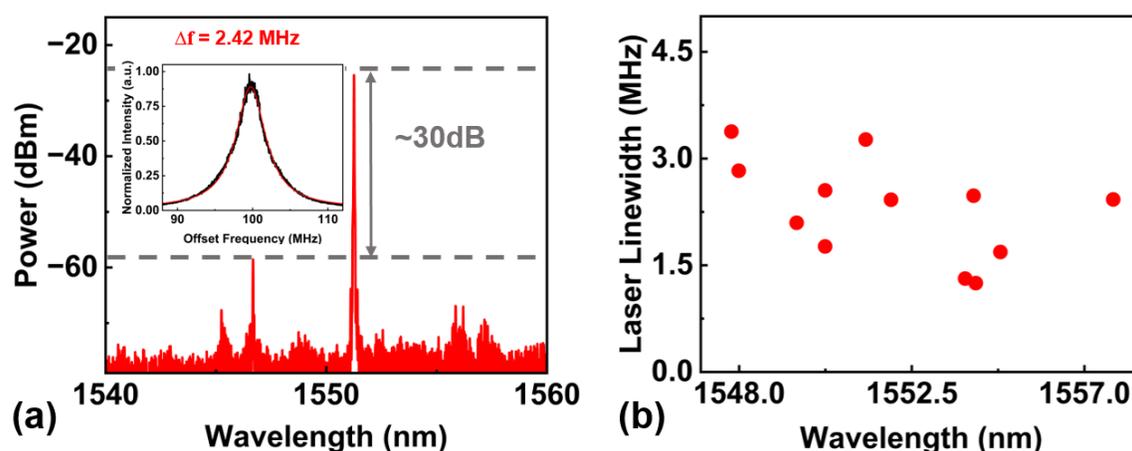

**Figure 5.** (a) A typical output spectrum from the laser at an injection current of 200 mA. (b) The intrinsic linewidth of the laser at different wavelengths..

As shown in Figure 5, the SMSR of the hybrid integrated tunable laser is 30dB, and the linewidth of the laser ranging from 1538nm to 1560nm is around 1-4 MHz. The above testing process is provided in the supplementary materials.

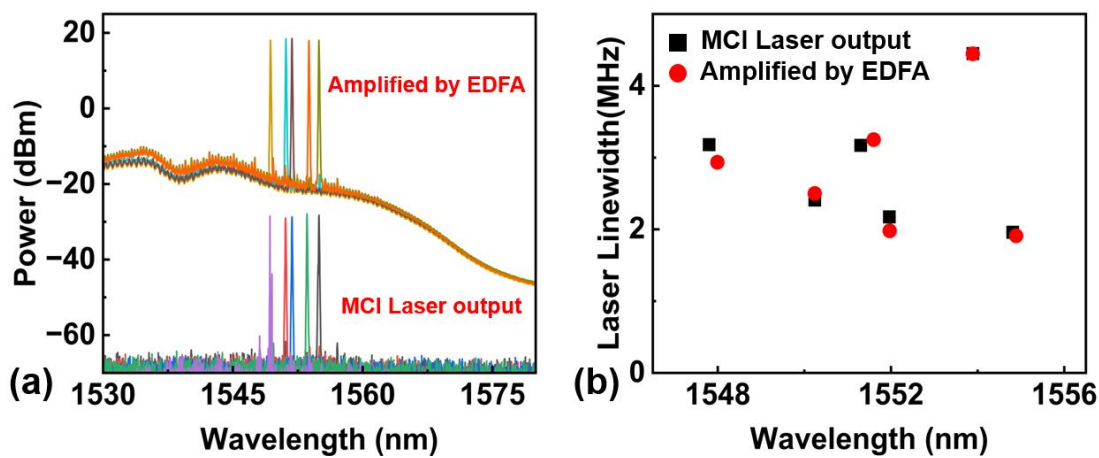

**Figure 6.** (a) Optical spectra of the amplified MCI laser and the MCI laser. (b) The intrinsic linewidth of the amplified MCI laser and the MCI laser.

After fiber coupling, the output power of the laser is about 10 μW, which can be further amplified to 70 mW by commercial amplifiers. The linewidth of the laser output after being amplified almost remain unchanged (see in Figure 6).

## 3   Discussion

In summary, by butt coupling a commercial RSOA with a lithium niobate-based MCI external cavity, we have demonstrated a hybrid integrated tunable laser. The laser optimizes the arm lengths of eight independent phase shifters to achieve longitudinal mode selection through MCI enhancement. Furthermore, the superior electro-optic tuning capabilities of lithium niobate allow for a wavelength tuning range of 22 nm which is accompanied by an excellent performance, characterized by a SMSR exceeding 30 dB and the feature of continuous tunability (with a tuning resolution of 0.014 nm). The tunable laser has the advantages of a simple structure, wide tuning range, and low power consumption. However, there is still room for further improvement in future research, mainly focusing on optimizing the electrical packaging of the laser. The wire bonding electrical packaging method can be replaced by better industrial-level packaging methods such as surface-mount technology, thereby fully utilizing the excellent electro-optic properties of lithium niobate material, and further improving the tuning speed, output characteristics, and integration of the chip. We look forward to these research efforts driving advancements in optical technology and playing a significant role in the fields of optical communication and other optical applications.

*Code and Data Availability*

Structural parameters, simulated and experimental data have been provided within the main text and Supplementary Material of this paper. All the other data that support the findings of this study are available from the corresponding authors upon reasonable request.

*Acknowledgments*

This work was supported by the National Key R&D Program of China (2019YFA0705000), Innovation Program for Quantum Science and Technology (2021ZD0301403), National Natural

Science Foundation of China (Grant Nos. 12334014,12192251), Science and Technology Commission of Shanghai Municipality(NO.21DZ1101500), Shanghai Municipal Science and Technology Major Project(2019SHZDZX01), supported by Fundamental Research Funds for the Central Universities